\documentclass[12pt, fleqn]{article}
\usepackage[cp1251]{inputenc}
\usepackage{latexsym,amsfonts,amssymb}
\usepackage{graphicx}

\usepackage{amsbsy}
\usepackage{amsmath}
\usepackage{epsf}

\newtheorem{theo}{Theorem}

\newcommand{\bt}{\begin{theo}}
\newcommand{\et}{\end{theo}}
\newcommand{\bd}{\begin{displaymath}}
\newcommand{\ed}{\end{displaymath}}

\newcommand{\be} {\begin{equation}}
\newcommand{\ee} {\end{equation}}
\newcommand{\ba}{\begin{array}{l}}
\newcommand{\ea} {\end{array}}
\newcommand{\bea}{\begin{eqnarray}}
\newcommand{\eea} {\end{eqnarray}}

\begin{document}

\begin{center}
 {\Large \textbf{ Comments on the paper `Derivation of lump solutions to a variety of Boussinesq
equations with distinct dimensions.'} }
\medskip\\
{\bf Roman Cherniha~$^{\dag,\dag\dag}$ \footnote{\small
Corresponding author. E-mails: r.m.cherniha@gmail.com;
roman.cherniha1@nottingham.ac.uk}}
 \\
{\it $^{\dag}$~Institute of Mathematics,  National Academy
of Sciences  of Ukraine,\\
 3, Tereshchenkivs'ka Street, Kyiv 01004, Ukraine \\
  $^{\dag\dag}$~School of Mathematical Sciences, University of Nottingham,\\
  University Park, Nottingham NG7 2RD, UK
}
 \end{center}

\begin{abstract}
The Comments are   devoted to the    paper  'Derivation of lump
solutions to a variety of Boussinesq equations with distinct
dimensions' (Int J Numer Methods Heat Fluid Flow.
2022;32:3072–3082), in which three new generalizations of the
classical Boussinesq equation are suggested that were further
investigated in several papers. Here it   is shown that the
equations derived in the above paper are not presented in their
canonical forms. It turns out that all three  equations can be
essentially simplified by the standard technique widely used for
linear and  quasi-linear PDEs. As a result, it is proved  that the
equations suggested are not multidimensional generalizations of the
Boussinesq equation.

\end{abstract}



The recent paper \cite{wazwaz-22}  and several others of the same
author are  devoted to study a variety of nonlinear equations that
are called Boussinesq equations in distinct dimensions. The author
consider those equations as non-trivial generalizations of the
classical Boussinesq equation
 \be \label {1} u_{tt}+u_{xx}-\beta(u^2)_{xx}-\gamma u_{xxxx}=
0,\ee
 where $u(t,x)$ is an unknown smooth function
  (the lower subscripts denote differentiation with
respect to relevant  variables in what follows).

A new integrable (1+1)-dimensional  Boussinesq equation is suggested
in the form \cite{wazwaz-22} \be \label {2}
u_{tt}+u_{xx}-\beta(u^2)_{xx}-\gamma u_{xxxx} +\alpha u_{xt}= 0\ee
(hereafter the parameters $\alpha, \beta, ...$ are nonzero
constants). However, if one applies  the well-known  technique used
for the reduction of   PDEs  to their canonical forms (this
technique is described in each textbook devoted to linear and
quasi-linear PDEs, see, for instance, the classical book
\cite{courant-hilbert}) then the PDE \be \label {2*} u_{t^*t^*}+(1-
\alpha^2/4)u_{x^*x^*}-\beta(u^2)_{x^*x^*}-\gamma u_{x^*x^*x^*x^*}=
0\ee is obtained by the transformation \be \label {3}t^*=t, \quad
x^*= x- \frac{\alpha}{2}t.  \ee
 Obviously, PDE (\ref{2*}) is nothing else
but the Boussinesq equation (\ref{1}) in new notations. The
coefficient $(1- \alpha^2/4)$ is reducible to 1 by the
transformation $\tau = \sqrt{1- \alpha^2/4}\ t^*, \ |\alpha|\not=2,$
while the second term simply vanish in the case $|\alpha|=2$.

A new  (1+2)-dimensional  Boussinesq equation is suggested in the
form \cite{wazwaz-22} \be \label {4}
u_{tt}+u_{xx}-\beta(u^2)_{xx}-\gamma u_{xxxx}+
\frac{\alpha^2}{4}u_{yy} +\alpha u_{yt}= 0.\ee The canonical form of
the above equation reads as \be \label {5}
u_{y^*y^*}+u_{xx}-\beta(u^2)_{xx}-\gamma u_{xxxx}= 0\ee and is
obtainable by the transformation \be \label
{6}t^*=t-\frac{2}{\alpha}\ y, \quad y^*= \frac{2}{\alpha}\ y.  \ee
Obviously, PDE (\ref{5}) is again the Boussinesq equation (\ref{1})
in new notations.

Finally, the so-called (1+3)-dimensional  Boussinesq equation is
proposed in the form \cite{wazwaz-22} \be \label {7}
u_{tt}+u_{xx}-\beta(u^2)_{xx}-\gamma u_{xxxx}+
\frac{\alpha^2}{4}u_{yy} +\alpha u_{yt}+ \delta u_{xz}= 0.\ee The
above PDE is reducible to the much simpler equation \be \label {8}
u_{y^*y^*}-\beta(u^2)_{x^*x^*}-\gamma u_{x^*x^*x^*x^*} +\delta
u_{x^*z} = 0\ee by the transformation \be \label
{9}t^*=t-\frac{2}{\alpha}\ y, \quad x^*= x- \frac{1}{\delta}z, \quad
y^*= \frac{2}{\alpha}\ y. \ee  It is very difficult to imagine that
PDE (\ref{8}) is  (1+3)-dimensional  Boussinesq equation if one
compares this equation and PDE (\ref{1}). On the other hand, one
easily notes that  PDE (\ref{8}) coincides  (up to notations and
parameter signs) with the classical  Kadomtsev–Petviashvili equation
 \be \label {10} u_{yy}- \lambda(u_{xt}+\frac{3}{2}(u^2)_{xx}+\gamma u_{xxxx})=
0.\ee So, PDE (\ref{7}) is equivalent to the KP equation and cannot
be called  the (1+3)-dimensional  Boussinesq equation.

Finally, it is a well-known fact that the  Boussinesq equation  and
the KP equation are integrable. So,  integrability and other
properties of the three equations investigated in \cite{wazwaz-22}
and several other papers
 are  trivial consequences of the integrability of the classical
equations (\ref{1}) and (\ref{10}).

\end{document}